\documentclass[
  ,final            
  ]
  {aipproc}

\layoutstyle{6x9}

\begin{document}

\begin{flushright}                                                   %
LU TP 03-33\\                                                        %
hep-ph/0307082\\July 2003                                            %
\end{flushright}                                                     %
\vfill                                                               %
\centerline{\bf                                                      %
CHIRAL DYNAMICS IN THE MESON SECTOR AT TWO LOOPS$^\dagger$}          %
\vfill                                                               %
\begin{center}                                                       %
{{\bf Johan Bijnens}\\[1cm]                                          %
Department of Theoretical Physics 2, Lund University,\\              %
S\"olvegatan 14A, S 22362 Lund, Sweden                               %
}                                                                    %
\end{center}                                                         %
\vfill                                                               %
\begin{center}{\bf Abstract}\end{center}                             %
I give a very short introduction to Chiral Perturbation Theory       %
and an overview of the next-to-next-to-leading order three-flavour   %
calculations done.                                                   %
I discuss those relevant                                             %
for an improvement in the accuracy of the measurement of $V_{us}$    %
in more detail. One major conclusion is that                         %
all needed $p^6$ low energy constants can be obtained                %
from experiment via the scalar form-factor in $K_{\ell3}$ decays.    %
\vfill                                                               %
$^\dagger$ Talk presented at CIPANP 2003, Conference on the          %
Intersections of Particle and Nuclear Physics,                       %
New York May 19-24, 2003.                                            %
\clearpage                                                           %

\title{Chiral Dynamics in the Meson Sector at two Loops}

\author{Johan Bijnens}{
  address={Department of Theoretical Physics 2, Lund University,\\
S\"olvegatan 14A, S 22362 Lund, Sweden}
}

\begin{abstract}
I give a very short introduction to Chiral Perturbation Theory
and an overview of the next-to-next-to-leading order three-flavour
calculations done.
I discuss those relevant
for an improvement in the accuracy of the measurement of $V_{us}$
in more detail. One major conclusion is that
all needed $p^6$ low energy constants can be obtained
from experiment via the scalar form-factor in $K_{\ell3}$ decays.
\end{abstract}

\maketitle


\def\be{\begin{equation}}
\def\ee{\end{equation}}
\def\ba{\begin{eqnarray}}
\def\ea{\end{eqnarray}}

\section{Introduction}
\label{intro}

The problem of dynamics of mesons at low energies is important.
It plays a major role in the precise determination of the elements of the
Cabibbo-Kobayashi-Maskawa matrix (CKM) which is a main part of the study
of the standard model flavour sector~\cite{CKM}.
In this talk I will concentrate on the theory behind the measurement
of $V_{us}$ from $K_{\ell3}$ ($K\to\pi\ell\nu$) decays and in particular
on the recent work of P.~Talavera and myself on the two-loop calculation
and the determination of the relevant low-energy constants~\cite{BT2,Moriond}.

I review shortly Chiral Perturbation Theory (ChPT) and
the relevant two-loop calculations done up to now, followed by
a discussion of $K_{\ell3}$ and the present theory situation. I
include a short discussion of the validity of the linear approximation of
the form factors normally used in the data analysis.
The main relevant results from ChPT~\cite{BT2} 
can be summarized as follows. The curvatures
are important in the analysis but can be predicted using ChPT from the pion
electromagnetic form-factor~\cite{BT2} and 
all order $p^6$ parameters needed to determine $V_{us}$
can be experimentally obtained via the scalar form factor, $f_0(t)$,
in $K_{\ell3}$~\cite{BT2}. The
curvature of $f_0(t)$ can be predicted as well from knowledge about
scalar form factors of
the pion~\cite{BD}, albeit only at fairly low precision at present.

\section{Chiral Perturbation Theory}

ChPT is an effective field theory valid as
an approximation to Quantum Chromodynamics (QCD) at low energies. 
Its modern form has was introduced by Weinberg,
Gasser and Leutwyler~\cite{ChPT,GL1}. 
The global chiral symmetry, $SU(3)_L\times SU(3)_R$, of QCD 
in the limit of
massless quarks is spontaneously broken down to the diagonal subgroup
$SU(3)_V$ by a nonzero quark condensate,
$
\langle \bar q q\rangle = \langle \bar q_L q_R+\bar q_R q_L\rangle
\ne 0\,.
$
The eight broken generators lead to eight Goldstone bosons. These are massless
{\em and} their interactions vanish at zero momentum. The latter allows
the construction of a well defined perturbative expansion in terms of momenta,
generically referred to as an expansion in $p^2$.
Quark masses are usually counted as order $p^2$ since
$p_\pi^2 = m_\pi^2 \sim m_q \langle\bar q q\rangle/F_\pi^2$. Inserting
an external photon or $W^\pm$-bosons counts as order $p$ since these
are included via covariant derivatives. Recent lectures, much more detailed
than what is included here are~\cite{chptlectures}.
ChPT being an effective field theory implies that
the number of parameters increases order by order. In the
purely mesonic strong and semi-leptonic sector there are two
parameters at lowest order ($p^2$), ten at NLO ($p^4$)~\cite{GL1}, and 90
at NNLO ($p^6$) \cite{BCE1}. The renormalization procedure and the divergences
are worked out in general to NNLO~\cite{BCE2} and provide a good check on
all calculations. One problem shared with other high order loop
calculations in comparing different calculations
is the use of different renormalization schemes.
The calculations that were used to determine all the needed parameters
are those of the masses and decay
constants \cite{ABT1}, $K_{\ell4}$~\cite{ABT2} and the
electromagnetic form factors~\cite{BT1}.

\section{$K_{\ell3}$: definitions, $V_{\lowercase{us}}$ and 
form factor linearity}

The neutral and charged $K_{\ell3}$ decays amplitudes,
$ K^{+,0}(p) \rightarrow \pi^{0,-} (p') \ell^+ (p_\ell) \nu _\ell (p_\nu)$,
are
\ba
 T^{(+,0)}
&=& \frac{G_F} {\sqrt{2}} V_{us}^\star \ell^\mu { F_\mu}^{(+,0)} (p',p)\,,
\quad\quad
\ell^\mu = \bar{u} (p_\nu)\gamma^\mu  (1- \gamma_5) v (p_\ell)\,,
\nonumber\\
{ F_\mu}^{+,0} (p',p)
&=& \left( {1}/{\sqrt{2}} , 1\right)\,
\left[(p'+p)_\mu f^{K^+\pi^0,K^0\pi^-}_+ (t) + (p-p')_\mu
f_-^{K^+\pi^0,K^0\pi^-} (t)\right]\!\!.
\ea
Isospin leads to the relations
\be
f^{K^0\pi^{-}}_+(t)=f^{K^+\pi^0}_+(t)= f_+(t)\quad\mbox{and}\quad
f^{K^0\pi^{-}}_- (t)=f^{K^+\pi^0}_- (t)= f_-(t)
\,,
\ee
The scalar form factor and the usual linear parameterizations are defined as
\be
 f_0 (t) = f_+ (t) + {t}/({m^2_K - m^2_\pi})\, f_-(t)\,,
\quad\quad
f_{+,0}(t) = 
f_+(0)\left(1+\lambda_{+,0}\,{t}/{m_\pi^2}\right)\,.
\ee
To determine $|V_{us}|$ we need $f_+(0)$ theoretically 
and experimentally. There are three
main theoretical effects. There is a well-known short-distance correction from
$G_\mu$ to $G_F$ calculated by Marciano and Sirlin. The corrections of
order $(m_s-\hat m)^2$ allowed by the Behrends-Sirlin-Ademollo-Gatto theorem 
are discussed at $p^6$ here and in \cite{BT2,Moriond}.
 The sizable isospin breaking found
by Leutwyler and Roos~\cite{LR} is in the process of being evaluated at order
$p^6$ too. On the experimental side, the old radiative correction calculations
used in~\cite{LR} are updated in~\cite{Radiative} where
a clean procedure with generalized form-factors is proposed.
The experimental data are mostly analyzed using a linear form factor
$f_+(t)$. The recent precise CPLEAR data~\cite{CPLEAR} allow
to test this assumption~\cite{BT2}. Using a linear fit to their data
and {\em neglecting} systematic errors we get a normalized $f_+(0)=1$
and $\lambda_+=0.0245\pm0.0006$. Allowing for curvature we obtain
a sizable curvature, $f_+(0) = 1.008\pm0.009$ and
$\lambda_+=0.0181\pm0.0068$. The fitted curvature is compatible with zero,
the central value is precisely at the ChPT prediction
given below. In order to obtain $|V_{us}|$ with an error of 1\% it is therefore
important to include the effect of curvature in the analysis.
The central value of $\lambda_+$ is outside the errors quoted for the
linear fit~\cite{CPLEAR}. After the meeting we obtained
similar conclusions for the KEK-PS E246 data~\cite{KEKPS},
see v2 of \cite{BT2}.

\section{$\lowercase{f_+(t)}$ and $\lowercase{f_0(t)}$: theory}

The  $f_+(t)$ ChPT calculation for is rather cumbersome.
I will use our work,~\cite{BT2} but an independent calculation
exists~\cite{Post3} and agrees reasonably
well, the difference is discussed in ~\cite{BT2}.
We write the amplitude as
\ba
f_+(t) &=& 1 + f_+^{(4)}(t)/F_\pi^2 + f_+^{(6)}(t)/F_\pi^4
\quad\mbox{with}\quad
f_+^{(4)}(t) = {t}{ L_9^r}/2+\mbox{loops}\,,
\nonumber\\
f_+^{(6)}(t) &=& -8 \left(C_{12}^r+C_{34}^r\right)
\left(m_K^2-m_\pi^2\right)^2
+ t\, { R_{+1}^{K\pi}}
+ t^2\, (  - 4 C_{88}^r + 4 C_{90}^r )
+\mbox{loops}(L_i^r)\,.
\ea
The pion electromagnetic form factor data yield~\cite{BT1}
$L_9^r = 0.00593\pm 0.00043$ and
$  - 4 C_{88}^r + 4 C_{90}^r = 0.00022\pm0.00002\,.$
With this input we fit the CPLEAR data and obtain
\begin{figure}
\includegraphics[width=8.5cm,angle=-90]{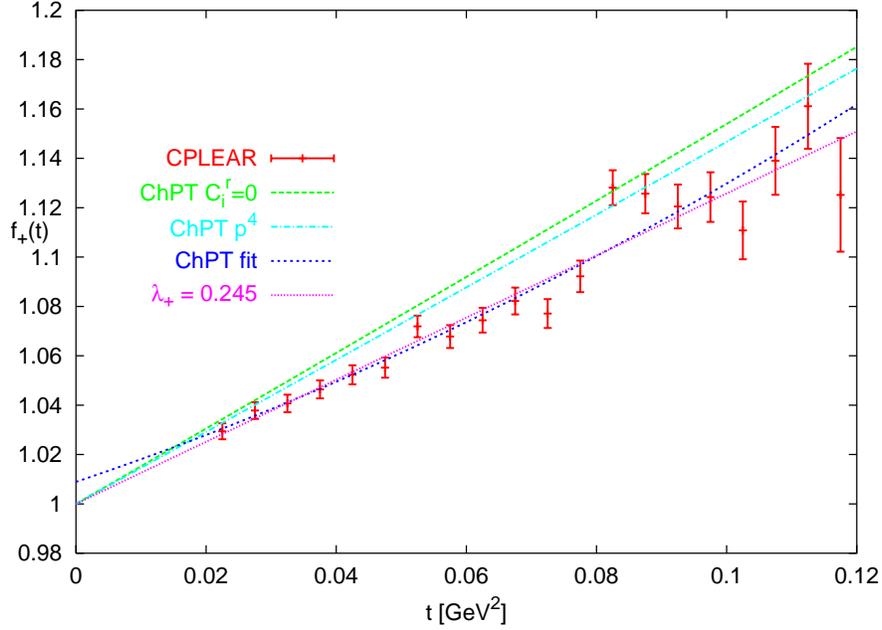}
\caption{ChPT fits to the CPLEAR data showing the effect of the
curvature compared to linear fits.
\label{fig:CPLEAR}}
\end{figure}
\be
{ R_{+1}^{K\pi}} =
 -(4.7\pm0.5)~10^{-5}~\mbox{GeV}^2\quad\mbox{and}\quad
 \lambda_+ = 0.0170\pm0.0015\,.
\ee
The first agrees with the VMD estimate~\cite{BT2}
 ${R_{+1}^{K\pi}|_{VMD}} \approx -4~10^{-5}~\mbox{GeV}^2$.
The latter comes from ChPT as 
\be
\lambda_+ = 0.0283~(p^4)~~ + 0.0011~(\mbox{loops }p^6)~~ -0.0124 (C_i^r)\,. 
\ee

Our main conclusion follows from rewriting the full
$p^6$ result for $f_0(t)$ as
\ba
f_0(t) &=& 1-({8}/{F_\pi^4})\,{\left(C_{12}^r+C_{34}^r\right)}
\left(m_K^2-m_\pi^2\right)^2
+(8 t/{F_\pi^4})\,{ \left(2C_{12}^r+C_{34}^r\right)}
\left(m_K^2+m_\pi^2\right)
\nonumber\\&&
+ t/\left(m_K^2-m_\pi^2\right)\,\left(F_K/F_\pi-1\right)
-(8 t^2/F_\pi^4)\, { C_{12}^r}
+\overline\Delta(t)+\Delta(0)\,.
\ea
{\em 
$\overline\Delta(t)$ and
$\Delta(0)$ contain  NO $C_i^r$ and only depend on the
$L_i^r$ at order $p^6$ thus ALL
needed parameters can be determined experimentally.}

$
\Delta(0) = -0.0080\pm0.0057[\mbox{loops}]\pm0.0028[L_i^r]\,,
$
is known and an expression for $\Delta(t)$ can be found in Ref.~\cite{BT2}
The errors are an estimate of higher orders and using fits of the $L_i^r$
using different assumptions. The $p^6$ estimate of
\cite{LR} corresponds to
$-({8}/{F_\pi^4})\,{\left(C_{12}^r+C_{34}^r\right)}
(m_K^2-m_\pi^2)^2 \approx -0.016\pm0.008$.

\section{Conclusions}

I discussed the $K_{\ell3}$
form factors in ChPT to order $p^6$.
The main conclusions are that the curvatures for $f_+(t)$ and $f_0(t)$
can be predicted from the data on pion electromagnetic~\cite{BT1}
 and scalar~\cite{BD}
form-factors, the curvature in $f_+(t)$ and $f_0(t)$
should be taken into account in new precision experiments but from the
slope and the curvature we can determine experimentally the needed
parameters to calculate $f_+(0)$. A precision of better than
one percent seems feasible for $|V_{us}|$. 


\begin{theacknowledgments} 
This work has been funded in part by
the Swedish Research Council and the European Union RTN
network, Contract No. HPRN-CT-2002-00311  (EURIDICE)
\end{theacknowledgments}



\begin{thebibliography}{99}
\bibitem{CKM} 
M.~Battaglia {\it et al.},
hep-ph/0304132.

\bibitem{BT2} J.~Bijnens and P.~Talavera,
hep-ph/0303103, to be published in Nucl. Phys. B.

\bibitem{Moriond}
J.~Bijnens,
Talk given at 38th Rencontres de Moriond on QCD and High-Energy Hadronic
Interactions, Les Arcs, Savoie, France, 22-29 Mar 2003,
hep-ph/0304284.

\bibitem{BD} J.~Bijnens and P.~Dhonte,
hep-ph/0307044

\bibitem{ChPT} S.~Weinberg,
Physica A {\bf 96} (1979) 327;
J.~Gasser and H.~Leutwyler,
Annals Phys.\  {\bf 158} (1984) 142,

\bibitem{GL1}
J.~Gasser and H.~Leutwyler,
Nucl.\ Phys.\ B {\bf 250} (1985) 465.


\bibitem{chptlectures}
A.~Pich, A., 
hep-ph/9806303;
G.~Ecker,
hep-ph/0011026;
S.~Scherer,
hep-ph/0210398.

\bibitem{BCE1}
J.~Bijnens, G.~Colangelo and G.~Ecker,
JHEP {\bf 9902} (1999) 020
[hep-ph/9902437];

\bibitem{BCE2}
J.~Bijnens, G.~Colangelo and G.~Ecker,
Annals Phys.\  {\bf 280} (2000) 100
[hep-ph/9907333].

\bibitem{ABT1}
G.~Amor\'os, J.~Bijnens and P.~Talavera,
Nucl.\ Phys.\ B {\bf 568} (2000) 319
[hep-ph/9907264],
Nucl.\ Phys.\ B {\bf 602} (2001) 87
[hep-ph/0101127].

\bibitem{ABT2}
G.~Amor\'os, J.~Bijnens and P.~Talavera,
Phys.\ Lett.\ B {\bf 480} (2000) 71
[hep-ph/9912398];
Nucl.\ Phys.\ B {\bf 585} (2000) 293
[Erratum-ibid.\ B {\bf 598} (2001) 665]
[hep-ph/0003258].


\bibitem{BT1}
J.~Bijnens and P.~Talavera,
JHEP {\bf 0203} (2002) 046
[hep-ph/0203049].

\bibitem{LR}
H.~Leutwyler and M.~Roos,
Z.\ Phys.\ C {\bf 25} (1984) 91.

\bibitem{Radiative}
V.~Cirigliano {\it et al.},
Eur.\ Phys.\ J.\ C {\bf 23} (2002) 121
[hep-ph/0110153].

\bibitem{CPLEAR}
A.~Apostolakis {\it et al.}  [CPLEAR Collaboration],
Phys.\ Lett.\ B {\bf 473} (2000) 186.

\bibitem{KEKPS}
A.~S.~Levchenko {\it et al.}  [KEK-PS E246 Collaboration],
Phys.\ Atom.\ Nucl.\  {\bf 65} (2002) 2232
[Yad.\ Fiz.\  {\bf 65} (2002) 2294]
[hep-ex/0111048].


\bibitem{Post3}
P.~Post and K.~Schilcher,
Eur.\ Phys.\ J.\ C {\bf 25} (2002) 427
[hep-ph/0112352].

\end{thebibliography}
\end{document}